# FLOODING ATTACKS TO INTERNET THREAT MONITORS (ITM): MODELING AND COUNTER MEASURES USING BOTNET AND HONEYPOTS

K.Munivara Prasad[1] and A.Rama Mohan Reddy[2]    M.Ganesh Karthik[3]

[1]Department of Computer Science and Engineering, Rayalaseema University, Kurnool
prasadkmv27@gmail.com

[2]Professor and Head,Departmentof Computer Science and Engineering,SVUCE,SV University,Tirupati
ramamohansvu@yahoo.com

[3]Department of Computer Science and Engineering, Sree vidyanikethan Engg.College,Tirupati
ganeshkarthik16@gmail.com

## ABSTRACT

*The Internet Threat Monitoring (ITM),is a globally scoped Internet monitoring system whose goal is to measure, detect, characterize, and track threats such as distribute denial of service(DDoS) attacks and worms. To block the monitoring system in the internet the attackers are targeted the ITM system. In this paper we address flooding attack against ITM system in which the attacker attempt to exhaust the network and ITM's resources, such as network bandwidth, computing power, or operating system data structures by sending the malicious traffic. We propose an information-theoretic frame work that models the flooding attacks using Botnet on ITM. Based on this model we generalize the flooding attacks and propose an effective attack detection using Honeypots.*

## 1. INTRODUCTION

The Internet was initially designed for openness and scalability. The infrastructure allows the user to misuse and it is the opportunity to the attackers there to perform some malicious transactions. On the Internet, anyone can send any packet to anyone without being authenticated, while the receiver has to process any packet that arrives to a provided service. The lack of authentication means that attackers can create a fake identity, and send malicious traffic with impunity. A denial-of-service (DoS) attack [2] is an explicit attempt by attackers to prevent an information service's legitimate users from using that service. These attacks, attempt to weaken and blocks the the victim's resources, such as bandwidth available in the network, computing power of the system, or operating system data structures. Flood attack, Ping of Death attack, SYN attack, Teardrop attack, DDoS, and Smurf attack are the most common types of DoS attacks. The hackers who launch DDoS attacks typically target sites or services provided by high-profile organizations, such as government agencies, banks, credit-card payment gateways, and even root name servers.

A flooding-based Distributed Denial of Service (DDoS) attack is performed by the attacker by sending a huge amount of unwanted traffic to the victim system and it is the very commonly used attack by the attacker. Network level congestion control can throttle peak traffic to protect the network. Network monitors are used to monitor the traffic in the networks to classify them as genuine or attack traffic and also these monitors gives the traffic as an input to several DDoS detection algorithms for detection of DDoS attacks. However, it cannot stop the quality of service (QoS) for legitimate traffic from going down because of attacks. Two features of DDoS attacks hinder the advancement of defense techniques. First, it is hard to distinguish between DDoS attack traffic and normal traffic. There is a lack of an effective differentiation mechanism that results in minimal collateral damage for legitimate traffic. Second, the sources of DDoS attacks are also difficult to find in a distributed environment. Therefore, it is difficult to stop a DDoS attack effectively.

The Internet Threat Monitoring (ITM) System basically has two main components one is centralized data center and another is the number of monitors which are distributed across the Internet. Each monitor covers the range of IP addresses and monitors the traffic to send the traffic logs to data center. The data center now collects the traffic logs from monitors and analyzes the collected traffic logs to publish reports to ITM system users.

The collected logs, as a random sample of the Internet traffic, can still provide critical insights for the public to measure, characterize, and track/detect Internet security threats. The idea of ITM systems dates back to DShield and CAIDA network telescope [4], [5][17], which have been successfully used to analyze the activities of worms and DDoS attacks [3], [6].

The reason is that if an attacker discovers the monitor locations, it can easily avoid detection (by ITM systems) by bypassing the monitored IP addresses and directing the attack to the much larger space of unmonitored IP addresses. Furthermore, such an attacker may even mislead the reports published by an ITM system by manipulating traffic to the identified monitors, generating highly skewed samples. Since ITM reports are trusted by the public as a random (unbiased) sample of Internet traffic, the confidentiality of monitor locations is vital for the usability of ITM systems.

The attacker compromises the monitor security and it's locations by using several attacks which includes Denial of service attacks (DoS), Distributed DoS (DDoS) and other attacks like Localization attacks [1]. The DoS and DDoS attacks creates the vulnerabilities or some loopholes in the software implementation to make that the resources available at the victim are bring down or blocked, which includes bandwidth attacks.

In this paper we introduce an information theoretic frame work to model existing flooding attacks in ITM system monitors. In the flooding attack the attacker sends the large volume of unwanted traffic to the targeted monitor for this he uses the botnet. Based on the Information-theoretic model we propose an effective approach to detect flooding attacks using Honey pots.

## 2. RELATED WORK

Probing traffic based Localization attack [7][8] in which an attacker sends high rate short length port scan messages to the targeted network to compromise the monitor locations in ITM system. Then, attacker queries the data center to determine whether a short spike of high-rate traffic appears in the queried time-series data, for confirmation of the attack.

A steganographic localization attack [9] an attacker launches a stream of low-rate port-scan probing traffic which is marginally modulated by a secret Pseudonoise (PN) code. While the low-rate property prevents the exhibition of obvious regularity of the published traffic data at the data center, based on the carefully synchronized PN code, the attacker can still accurately identify the PN-code-modulated traffic in the retrieved published traffic data from the data center. Thereby, the existence of monitors in the targeted network can be compromised. To this end, the PN-code-based steganographic attack presented in our paper can be understood as a covert channel problem [10], because the attack traffic encoded by a signal blends into the background traffic and is only recognizable by the attacker which knows the secret pattern of the PN code.

In [1] introduced the information theoretic framework to evaluate the effectiveness of the localization attacks by using the minimum time length required by an attacker to achieve a predefined detection rate as the metric. But this frame work is defined in specific to the localization attacks only; they are not given any solution for other DDoS attacks. The frame work allows the ITMs which are registered within the data center given, and the access is restricted to that private region only. But public access of the ITMs and data center allows more scope to provide security against different attacks.

## 3. PROPOSED WORK

In [1] the authors define a model in which the ITMs in the networks sends the traffic logs periodically to the data center and the data center collects the traffic logs and publishes the reports to ITM system users which are registered, that means it creates the private environment or region .In the private region the scope for DDoS attacks are very less, and they are restricted this model only for Localization attacks. In this section we have defined a model which will provide the following extensions.

***Public accessing:*** Public accessing of the data center increases the network usage and provides better communication with the outside world rather than private environment. In this any user from outside the private region can get the communication with the private network, if the user is genuine he can get the status of the monitor before sending the data to internal monitors, to avoid the attacks. If the user is an attacker, then this status information can be misused to perform the attacks on the monitor. The data center sends the status information to any users (public or private) based on the request query, but the private (internal) users can get the highest priority.

**Usage of Botnets for Flooding Attack:**

A denial-of-service (DoS) attack is an explicit attempt by attackers to prevent an information service's legitimate users from using that service. In a DDoS attack, these attempts come from a large number of distributed hosts that coordinate to flood the victim with an abundance of attack packets simultaneously. The attacker may use the botnets [11], [12] and other alternatives to launch the attack.

### 3.1 Flooding:

*Launching a flooding attack:* Once the DDoS network has been set up and the infrastructure for communication between the agents and the handlers established, all that an attacker needs to do is to issue commands to the agents to start sending packets to the victim host. The agents try to send unusual data packets (all TCP flags set, repeated TCP SYN packets, Large ICMP packets) to maximize the possibility of causing disruption at the victim and the intermediate nodes. There are certain basic packet attack types which are favorites of the attack tool designers. All the attack tools use a combination of these packet attack types to launch a DDoS attack. The basic attack types are

*i) TCP floods:* A stream of packets with various flags (SYN,RST, ACK) are sent to the victim machine. The TCP SYN flood works by exhausting the TCP connection queue of the host and thus denying legitimate connection requests. TCP ACK floods can cause disruption at the nodes corresponding to the host addresses of the floods as well. Also the one known tool that uses TCP ACK flooding (mstream [13]) has been known to cause disruptions in a router even with a moderate packet rate. Both TCP SYN flooding and the mstream attack constitute a group of attacks known as asymmetric attacks (Attacks where a less powerful system can render a much more powerful system useless).

*ii) ICMP floods (e.g ping floods):* A stream of ICMP packets is sent to the victim host. A variant of the ICMP floods is the Smurf attack in which a spoofed IP packet consisting of an ICMP ECHO_REQUEST is sent to a directed broadcast address. The RFC for ICMP specifies that no ECHO_REPLY packets should be generated for broadcast addresses, but unfortunately many operating systems and router vendors have failed to incorporate this into their implementations. As a result, the victim host (in this case the machine whose IP address was spoofed by the attacker) receives ICMP ECHO_REPLY packets from all the hosts on the network and can easily crash under such loads. Such networks are known as amplifier networks and thousands of such networks have been documented.

*iii) UDP floods:* A huge amount of UDP packets are sent to the victim host. Trinoo is a popular DDoS tool that uses UDP floods as one of its attack payloads.

## 3.2 BOTs

The attacker uses the bots to generate huge number of packets to attack the victim by sending these huge packets as large traffic to generate flooding attack. The attacker first identifies the compromised servers in terms of security and controls the systems which are under the control of these compromised servers. The compromised systems under the servers known as the bot. Normally the attacker communicates the bots by using the Internet relay chat (IRC)[14] .IRC is the public network there the users can enter and communicate each other or with  the groups openly.

 The attacker launches the DDoS attacks through the bots by sending the commands using these IRC network. The DDoS attacks can be blocked or the detection can not be possible, but by identifying the IRC server one can block the packets to the victim.

## 3.3 Botnet:

          Internet is the globally established network where different users or systems exist and it provides the better scalability and openness to the users in terms of services. The open accessing of the internet allows different threats and one of the major threats is from large number of compromised computers also called as bots or Zombies and the group of these computers called as Botnet. By using these botnets the attackers performs the attacks on the victims by simply sitting in house, from offices or organizations and any private or public network around the world. Every botnet or the group of compromised bots is controlled by a master commonly called as attacker or hacker. These botnets conducts various attacks which includes DDoS ,e-mail spamming, keylogging, click fraud, and spreading any malware to the victim. Compared to any attack the botnets consists of pool of compromised bots and these are capable to conduct or   damage the victim tremendously with collective power or capacity than the individual attacker. Example for these type of attacks are flooding, flash crowd and ports scan attacks there the attacker uses the botnet power to generate the large number of traffic to blocks the victim resources.

 **Attacking Behavior :** During the preparation an attack, botnets normally generate a large amount of malicious traffic, which in turn can make possible of easy detection. Understanding the attacks requires the attacking behaviour and a lot more information can reveal important intelligence including the nature of botnet, purpose of hackers and the origin of hackers. The attacking behaviours can defined from the following four aspects:

- Infecting new hosts
- Stealing personal information
- Phishing and spam proxy
- DDoS

**Infecting new hosts:** Botnets often selects new hosts using same ways as the virus and worms do for attack. The method that botnets use to compromise new hosts is through social engineering and distribution of malicious emails. In general the botnet distributes the malware using the mail attachments. Social engineering techniques are used to trap computer users into executing the malware, which leads to the compromise of hosts.

**Stealing Sensitive Information:** Recent botnets have employed complicated tools to steal sensitive user information from compromised hosts. The most commonly used tools for stealing sensitive information at victim systems are keyloggers and network traffic sniffers. Keyloggers modify host operating systems to spy on user activities and store user key strikes**.** Network traffic sniffers monitor network traffic sent over the subnet of the compromised host. The sensitive data is logged by these tools and then compiled into digested formats. Periodically, the data will be sent to their bot masters using various communication channels. Some commonly used methods are to send data through a designated IRC channel created by a botnet and in emails to a designated email address. BKDR_WAR.B steals keystrokes on a compromised computer in this way.

**Sending Spam:** Botnets are widely used to broadcast spam for different attack purposes. Two major advantages for hackers to use botnets to distribute spam are that the victims cannot trace the spam back to the source for legal action, and botnets can distribute a much larger volume of spam because of the aggregate computing power and vast availability of bandwidth. While some spam is used to distribute exploits (malware) as described in a previous subsection, some spam tricks users into visiting certain malicious websites, which install malware on their computers by exploiting Internet browser vulnerabilities.

**Distributed Denial of Service:** A DDoS attack [19] is probably one of the oldest botnet attack mechanisms. In the infancy of botnets, hackers began using botnets to launch DDoS attacks against a number of large organizations to consume all of their available platform CPU cycles and available bandwidth, effectively slowing their services down to a crawl, or knocking out their services altogether. For example, both Yahoo! and Microsoft were victimized by DDoS attacks launched by botnets in the past years. DDoS attacks still occur, but in a lesser frequency and volume. DDoS attacks have even recently been used for extortion. Botnets usually integrate a large variety attacking tools (e.g., UDP flooding, TCP SYN flooding, HTTP flooding). Some bots, such as PhatBot , even have very customized DDoS tools integrated into their code. AgoBot, SDBot, PhatBot, and many other botnets are all capable of launching DDoS attacks against a variety of targets.

**Botnet Life Cycle:** The botnet life cycle specifies that how these botnet are organised,planned in formation ,generation and propagation. The botnet has the following phases:

**1.** Bot-herder configures initial bot parameters.
**2.** Registers a DDNS.
**3.** Register a static IP.

**4.** Way of infecting the victim machines either directly through network or indirectly through user interaction.

**5.** Bots spread.

6. Adding the bots to Botnet through C & C Server.

7. Bots are used for some activity (DDoS, Identity Theft etc.)

8. Bots are updated through their Botoperator which issues update commands.

### 3.4 IRC-based Command and Control

A bot generally communicates with a controller to receive commands send by the attacker or send back information any to the attacker. It establishes the communication channel directly to the controller for transactions. The problem is that this connection could compromise the controller's location. Instead, the bot controller can use a proxy such as public message drop point (e.g., a well-known message board). The websites and other drop points can introduce significant communication latency; a more active approach is attractive. The commonly used communication channel is through the IRC.

IRC provides a common protocol that is widely used across the Internet and has simple text-based command syntax. There is also a large number of existing IRC networks that can be used as public exchange points. In addition, most IRC networks lack any strong authentication, and a number of tools to provide anonymity on IRC networks are available. Thus, IRC provides a simple, low-latency, widely available, and anonymous command and control channel for botnet communication. An IRC network is composed of one or more IRC servers as depicted in figure 1.

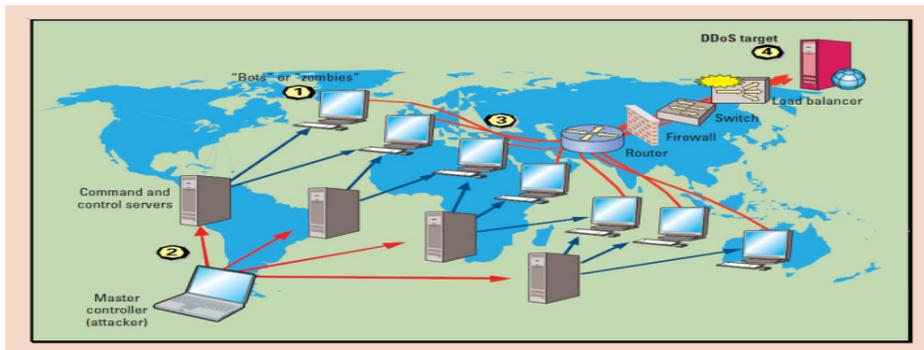

Figure 1: Compromised computers. In a distributed denial-of-service attack (DDoS), these computers serve three major roles: master controller, command and control server, and bot.

In the botnet communication every bot connects to a public IRC network or a hidden IRC serveron the compromised system.The bot receives the commands directly from the IRC controller by entering the named channel.

### 3.5 Honeypots:

A honeypot is an effective tool for observing and understanding the behaviour of intruder's tactics and intensions. A honeypot checks every packet transmitted to/from it, giving it the ability to collect highly determined and less noisy datasets for network attack analysis. Honeypots are trick computer resources set up for the purpose of monitoring and logging the activities of entities that probe, attack or compromise them. Honeypots can have many shapes and sizes; examples include dummy items in a database, low-interaction network components like preconfigured traffic sinks, or full-interaction hosts with real operating systems and services.

Honeypots reduces the burden on the servers or the systems interms of detection and logging. By capturing the small data sets of high volume it reduces the false positives and also captures the unknown attacks.

**Honeypot-based Research:**

Often we have a lack of precise information regarding attacks on the Internet. In most cases, we just see the results of attacks against networks or specific computers. For example, after a successful attack we just see that the compromised computer attacks further computers within the network. But analization of attacker strategy is very difficult to find. In addition to the problems gathering qualitative data, there is relatively little quantitative data on attacks against computer systems. In general the tools, strategies, and motives involved in computer and network attacks are difficult to define.

The term honeypot[16] generally refers to an entity with certain features that make it especially attractive and can tempt attackers into its vicinity. Honeypots are electronic tease, i.e. network resources (computers, routers, switches, etc.) deployed to be probed, attacked, and compromised. So called low-interaction honeypots simulate certain services – often on a massively parallel scale. High-interactions honeypots are real world systems which run special software permanently collecting data about the system and greatly aiding postincident computer and network forensics. A honeypot is usually a computer system with no production tasks in the network. This aids in detection of incidents: Every interaction with the system is suspicious and could point to a possibly malicious action. The potential for a zero false-positives rate is a clear advantage of honeypots in contrast to intrusion detection systems (IDS). Because of the wealth of data collected through them, honeynets are considered a very useful tool to learn more about attack patterns and attacker behavior in communication networks like the Internet.

Therefore honeypots have a dual use: They can be used as research instrument to learn about the tools, tactics, and techniques of intruders into IT-systems. In this usage scenario, qualitative and quantitative research is possible. Usually high-interaction honeypots tend to be used for qualitative research while low-interaction honeypots seem to be better suited for quantitative research.

# 4. PROPOSED MODEL

In this paper we divided the entire model into two regions namely private region and public region. The Internet Threat Monitors (ITM) are distributed across the Internet and each monitor records the traffic addressed to range of IP addresses and send the traffic logs periodically to the data center. The data center then analyzes the traffic logs collected from the monitors and publishes the reports to ITM system users. The collection of monitors under the data center forms the private region because the ITMs are registered before sending the logs to the data center. Any user can get the reports of the requested ITM by sending the query request to the data center and the data center is answerable to all the ITMs which are registered.

The public region of our model specifies the unregistered users of the data center who does not have any permission to access the data center, but they can get the traffic reports related to any ITM by sending the query request to the data center. The data center scope is extended to the public domain but it can only give the traffic reports to the public users. Allowing the public users or network accessing to the data center and monitors, causes decrease in the performance because of the overload of the data center. These can be balanced by introducing the priorities to the users; the internal or private region users have the highest priority than the public users .This priorities does not disturb the existing scenario but this can enhance the service to the public domain ,this will not be a over burden to the data center..

In This section we are constructing the botnet as the public user network without having any registration with data center and performing the flooding attack on the ITM which is local to the data center.

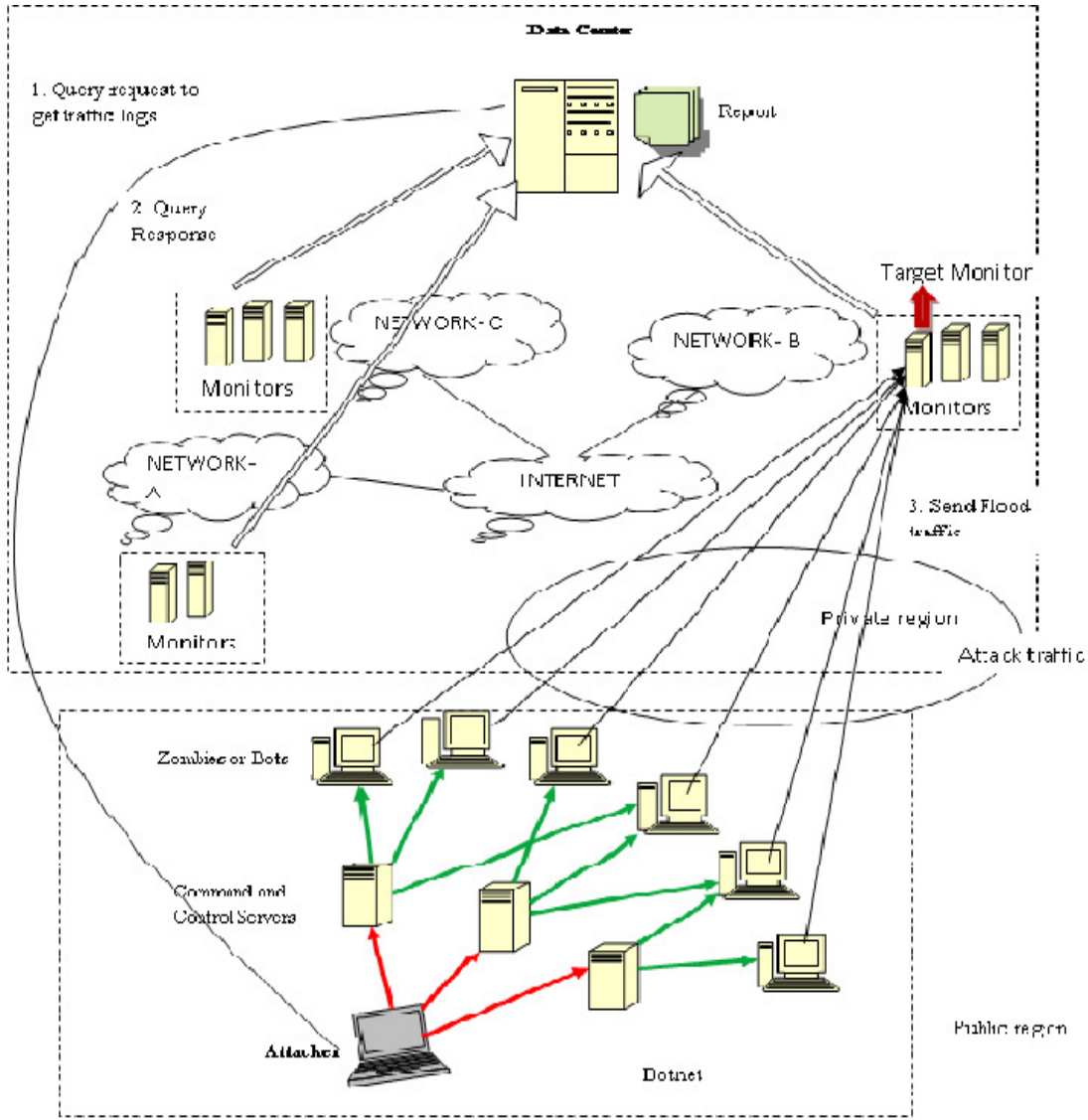

Figure 2: Work flow of flooding attacks using botnet.

**Generation of flooding attack with Botnet:**

A DDoS (Flooding) attack mechanism typically includes a network of several compromised computers [15]. These compromised computers serve three major role -master controller, command and control (C&C) server, and bot. An attacker generates a DDoS attack by exploiting vulnerabilities in one computer system and making it the DDoS "master controller." From here, the attacker identifies and communicates with other compromised systems. A C&C server is a compromised host with a special program running on it, this server distributes instructions from the attacker to the rest of the bots, which form a botnet[11]. (A bot is a compromised host that runs a special program.) Each C&C server is capable of controlling multiple bots, each of which is responsible for generating a stream of packets to the intended victim. Often, the bots employed to send the flood of requests are infected with a virus that lets attackers use them anonymously.

A Flooding attack happens in several phases:

• *Discover vulnerable hosts*. To launch a DDoS attack, attackers first build a network of computers that they can use to produce the volume of traffic needed to deny services to legitimate users. To create this network, they first scan and identify vulnerable sites or hosts. Vulnerable hosts are the systems usually those that run either no antivirus software or an out-of-date version. Attackers use these compromised hosts for further scanning and compromises.

• *Establish a botnet*. After gaining access, attacker must then install attack tools on the compromised hosts to form a botnet.

• *Launch an attack*. In the next phase, attackers send commands to C&C servers for their bots to attack by sending hundreds of thousands of requests to the target simultaneously.

• *Flood a target*. In the final phase, monitor receives a flood of requests to the point where they can't operate effectively.

# 5. PREVENTION

Preventive mechanisms attempt either to reduce the possibility of DDoS attacks or enable potential victims to endure the attack without denying services to legitimate users.

• *System security mechanisms* increase a host's overall security posture and prevent it from becoming part of a botnet or a DDoS victim. Examples of system security mechanisms include reliable firewall filtering, proper system configuration, effective vulnerability management, timely patch installation, robust antivirus programs, controlled and monitored system access, and solid instruction detection.

• *Resource multiplication mechanisms* provide an abundance of resources to counter DDoS threats, such as increasing the capacity of network bandwidth, routers, firewalls, and servers. Additional examples include deploying information services at diverse network locations and establishing clusters of servers with load-balancing capabilities. Resource multiplication essentially raises the bar on how many bots must participate in an attack to be effective. While not providing perfect protection, this last approach has often proved sufficient for small- to mid-range DDoS attacks.

**Preventing Flooding Attacks**

In this section we introduce a general methodology to prevent flooding attacks. It is based on the following line of reasoning:

1. To mount a successful Flooding attack, a large number of compromised machines are necessary.

2. To coordinate a large number of machines, the attacker needs a remote control mechanism.

3. If the remote control mechanism is disabled, the Flooding attack is prevented.

Our methodology to mitigate flooding attacks aims at manipulating the root-cause of the attacks, i.e., influencing the remote control network. Our approach is based on three steps:

1. Infiltrating the remote control network.

2. Analyzing the network in detail.

3. Shutting down the remote control network.

In the first step, we have to identify a method to smuggle an agent into the control network. In this context, the term agent describes a general procedure to mask as a valid member of the control network. This agent must thus be customized to the type of network we want to plant it in. The level of adaptation to a real member of the network depends on the target we want to infiltrate. For instance, to infiltrate a botnet we would try to simulate a valid bot, maybe even emulating some bot commands.

Once we are able to sneak an agent into the remote control network, it enables us to perform the second step, i.e., to observe the network in detail. So we can start to monitor all activity and analyze all information we have collected.

In the last step, we use the collected information to shut down the remote control network. Once this is done, we have deprived the attacker's control over the other machines and thus efficiently stopped the threat of a flooding attack with this network. Again, the particular way in which the network is shut down depends on the type of network.

# 6. DETECTION OF FLOODING ATTACKS

In this section we present efficient way of detecting the attacks on the ITMs in the given information theoretic frame work. We divide the attack detection process into two phases, Firstly the primary detection of DDoS attacks [20] on the ITMs and the later is the detection of flooding attacks on the ITMs.

In the primary detection phases the system detects the attacks based on traffic information aggregated from all monitors in the ITM system. If the overall traffic rate (e.g., volume in a given time interval) exceeds a predetermined threshold, the defender issues an alarm. The threshold value can be maintained either at data center or the individual ITMs based on the type of schemes used [1] in the network.In the primary detection phase the system detects some attack was happened in the network. If the detection scheme is centralized, then whenever the aggregate traffic exceeds the threshold maintained at the data center then the data center finds the attack and that attacked monitor can be identified by verifying the individual traffic logs of each ITM from the report. Otherwise if the detection strategy is distributed then each monitor maintained an individual threshold and checked the aggregate traffic regularly. If the traffic exceeds the threshold then it find the attack was happened and sends the status as attacked to the data center. After getting the attacked status from the ITM the data center blocks the corresponding ITM and displays the status of the ITM as blocked in the status reports, which will avoids the further traffic to or from the attacked ITM with the rest of the networks.

The second stage of detection specifies the detection of the flooding attacks. Once the attack is conformed then the data center identifies the attacked monitor and the traffic logs will be handover to the flooding detection phase. In this paper the flooding attacks are generated using botnet, so botnet tracking is required to detect and block the flooding attacks on the attacked ITM.

In this section we define the approaches for detecting the botnet. Once the botnet is successfully identified and blocked then automatically the flooding attacks can be avoided. In this connection the honeypots play the major role to block the botnet by identifying the command and control through the IRC server.

### 6.1 BOTNET Detection

Botnets are a very real and quickly evolving problem that is still not well understood. In this paer we discussed about the way hoe the bots are detected and stopped. We identify three approaches for handling botnets:

(1) Prevent systems from being infected,

(2) Directly detect command and control communication among bots and between bots and controllers, and,

(3) Detect the secondary features of a bot infection which includes propagation or attacks.

The first approach is to prevent the system from the attack, these can be done by using by introducing the anti-virus software, firewall or any security measures in the system.

The detection of command and control defines the second approach .The controlling of botnets are done in general with IRC and detection of IRC can be done by monitoring the TCP port 6667 which is used for IRC traffic. . One could also look for non-human behavioural characteristics in traffic, or even build IRC server scanners to identify potential botnets.

The third approach used to detect the botnet is purely depends on the identification of secondary features of bot infection or attack behaviour. Finding the command and control directly is not possible in this approach this can be done based on the the correlation of data from different sources to locate bots.

In this paper we explore the second and third approach for stopping botnets. The problem with the first approach is that preventing all systems on the Internet from being infected is nearly an impossible challenge. As a result, there will be large pools of vulnerable systems connected to the Internet for many years to come.

## 6.2 Detecting Command and Control

To avoid the damage of bots, we identified two approaches for detecting botnets: detect the command and control communication, or detect the secondary features of a bot infection.

### IRC-based Botnet Detection

Today, most known bots use IRC as a communication protocol, and there are several characteristics of IRC that can be leveraged to detect bots. One of the simplest methods of detecting IRC-based botnets is to offramp traffic from a live network on known IRC ports (e.g., TCP port 6667) and then inspects the payloads for strings that match known botnet commands. Unfortunately, botnets can run on non-standard ports. Another method is to look for behavioral characteristics of bots. One study found that bots on IRC were idle most of the time and would respond faster than a human upon receiving a command. The system they designed looked for these characteristics in Netflow traffic and attempted to tag certain connections as potential bots [15].

The idle IRC activity was successfully detected using this method but it is unable to find high false positive rate, for this honeypots is used to minimise the false positives.

One attack pool set up a vulnerable system and waited for it to be infected with a bot. They then located outgoing connections to IRC networks and used their own bot to connect back and profile the IRC server [16].

Honeypots are used to connect the bots directly rather than connecting IRC server and these honeypot checks the characteristics of command and control in outgoing connections. We identified all successful outgoing TCP connections and verified that they were all directly related to command and control activity by checking the payloads. There were a wide range of interesting behaviors, including connections from the bot to search engines to locate and use bandwidth testers, downloading posts from popular message boards to get server addresses, and the transmission of comprehensive host profiles to other servers.

These profiles consists of detailed information about the operating system, host bandwidth, users, passwords of the users, file shares, filenames and permissions for all files in the system, and a number of other minute details about the infected host. We then analyzed all successful outgoing connections for specific characteristics that could be used to identify botnet command and control traffic.

## 6.3 Collecting Malware with Honeypots

A honeypot is a network resource (computers, routers, switches, etc.) deployed to be probed, attacked, and compromised. A honeynet is a network of honeypots. Honeypot is a special software which periodically collects data about the system behaviour and provides automatic post-incident forensic analysis. The collected data enables us to determine the necessary information about an existing botnet.

The honeypots collets the data or attack traffic either from the data center or ITMs based on the detection scheme .Two types of detection schemes are defined in this paper based on the position of the honeypots.

*Centralized scheme:* In this approach only one honeypot is used for the detection and it is placed at the data center. Once the data center identifies the attacked ITM , then the traffic logs of the attacked ITM are send to the centralized honeypot to find the botnet.

*Distributed Scheme:* In the distributed approach the honeypots are placed at each ITM of the data center. Whenever the data center identifies the attacked monitor either by using centralized or distributed threshold detection approaches, then the attack traffic can be handover to the attached honeypot of that attacked ITM. The honeypot then identifies the botnet which causes the flooding attack.

The centralized scheme is more economical when compared to the distributed scheme because it uses only one honeypot at the data center instead of using individual honeypots for each ITM.But the efficiency of the system depends on the number of honeypots used in the network, If honeypots are more in the network then the detection of botnets is very simple and easy. whenever more than one ITMs are attacked in the network ,then the centralized scheme is less efficient than the distributed scheme.

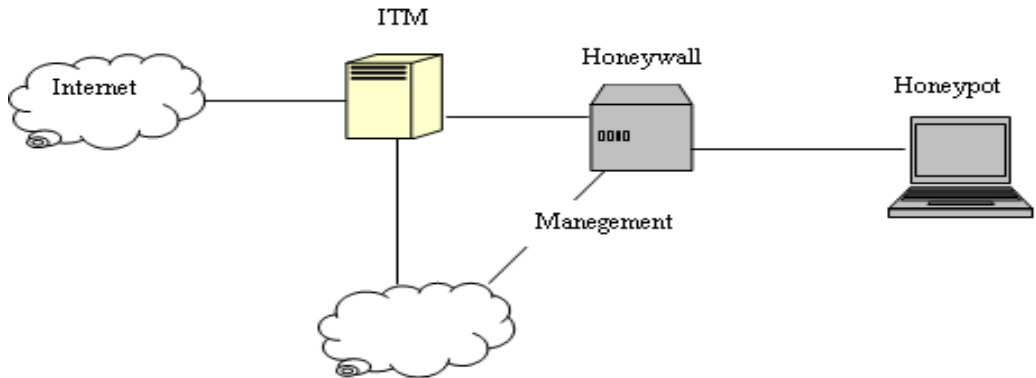

**Figure 3:Set up for tracking botnet using Honeypot**

GenII Honetpot is a windows honeypoy and used to collect the necessary information about the attack. The Windows honeypot runs an unwatched version of Windows 2000 or Windows XP. This system is thus very vulnerable to perform the attacks. The average expected lifespan of the honeypot is less than ten minutes. The shortest compromise time was only a few seconds: Once we plugged the network cable in, a bot compromised the machine and installed itself on the machine.

As explained in the previous section, a bot tries to connect to the C&C server to obtain further commands once it successfully attacked the honeypot. This is where the Honeywall comes into play. The Honeywall is a transparent bridge that enables the two tasks Data Control and Data Capture. Due to the Data Control facilities, it is possible to control the outgoing traffic. Using available tools for Data Control we can replace all suspicious in- and outgoing messages. A message is suspicious if it contains typical IRC messages for command and control, for example " TOPIC ", " PRIVMSG ", or " NOTICE ". Thus we are able to reduce the bot from accepting valid commands from the master channel. It can therefore cause no harm to others and therefore we have caught a bot inside our Honeynet. In addition with the detection, we can also derive all necessary sensitive information for a botnet from the data we have obtained up to that point in time: The Data Capture capability of the Honeywall allows us to determine the DNS/IP address of the bot which wants to connet the IRC.

In addition, we can obtain from the Data Capture logs the nickname, the indent information, the server's password, channel name, and the channel password as well. So we have collected all necessary information about the attack and the honeypot can catch further malware. Since we do not care about the captured malware for now, we rebuild the honeypot every 24 hours to have a "clean" system every day. This 10 has proven to be a good time span since after this amount of time the honeypot tends to become unstable due to installed malware.

# 7. CONCLUSION AND FUTURE WORK

The approach integrates active real time flooding attack flow identification from botnet with determining required number of honeypots. The honeypot controller has been modeled at Data center or ITMs to trigger honeypot generation in response to suspected attacks and route the attack traffic to honeypots. The performance of the proposed scheme is independent of attack traffic due to presence of honeypots at data center or ITMs. It gives stable network functionality even in the presence of high attack load.

Some of the avenues for further extensions are with larger and heterogeneous networks. Back tracking can be applied on attack flows to reach the attack source. Both of them hold promise for evaluating and improving our DDoS detection and defense method and data center information protection. The data center load can be still minimized by used some distributed load sharing algorithms.